\documentclass[12pt,a4paper,twoside]{article}  
\usepackage[cp1251]{inputenc}
\usepackage[T1,T2A]{fontenc}
\usepackage{amsmath,amsfonts,amssymb,amscd,euscript}
\usepackage[english]{babel}
\usepackage{url}
\usepackage[implicit=false]{hyperref}

\usepackage{graphicx}
\usepackage{psfrag}
\usepackage{cite}
\usepackage{caption}
\usepackage[textwidth=166mm,top=2cm,textheight=250mm,left=20mm,showframe=false]{geometry}
\usepackage{ifthen}

\makeatletter

\def\annotationandkeywordseng{\begin{otherlanguage}{english}\noindent {\small \referateng \par } \vspace{8pt}
\noindent {\small {\it Keywords}: \keywordseng} \par%
\vspace{8pt}
\noindent {\small {\textrm DOI:} \paperdoi} \par%
\vspace{10pt plus 6pt minus 1pt}\end{otherlanguage}}

\def\titleeng{\begin{otherlanguage}{english}\thispagestyle{firstpagestyleeng}\label{paperfirstpage}
\hbox{MSC2010: \MSC}
\vspace{30pt plus 6pt}
\begin{flushleft}
{\bf\copyright~{\textit{\authorseng}}\\[2ex]
{\MakeUppercase{\articletitleeng}}}
\end{flushleft}
\end{otherlanguage}}

\def\refereng{\begin{otherlanguage}{english}\vspace{20pt}  \par \small  \noindent \referateng\par\vspace{3ex}\@fundingeng\par\end{otherlanguage}}

\def\receivedeng{\begin{otherlanguage}{english}\vspace{3ex} \hfill Received~~\datereceive \par \vspace{5ex} \par\end{otherlanguage}}

\def\contactseng{\begin{otherlanguage}{english}\noindent\parbox{\textwidth}{\small\noindent\contactinformationeng\par\vspace{10pt}\par
\noindent{\bf Citation: }\authorseng. \articletitleeng, {\it Vestnik Udmurtskogo Universiteta. Matematika. Mekhanika. Komp'yuternye Nauki}, \paperyear, vol.~\papervolume, issue~\papernumber, \mbox{pp.~132--148.} \mbox{\url{https://doi.org/10.35634/vm210110}}}\label{paperlastpage}\end{otherlanguage}}

\@addtoreset{equation}{section}
\renewcommand{\section}{\@startsection{section}{1}{0pt}{1.3ex
plus 1ex minus .1ex}{1.3ex plus .1ex}{\bf\,\S\,}}
\newcommand{\point}{\hspace*{-4mm}{\bf.}\;}
\newcommand{\sect}[1]{\begin{flushleft}%
\vspace{8pt}
\protect{\section{\texorpdfstring{\point#1}{#1}}}
\vspace{8pt}
\end{flushleft}}

\renewcommand{\@begintheorem}[2]{\begin{trivlist}
\item[\hspace{\labelsep}{\bf \mbox{~~~}#1\ #2.}]}
\renewcommand{\@opargbegintheorem}[3]{\begin{trivlist}
\item[\hspace{\labelsep}{\bf \mbox{~~~}#1\ #2 {\rm (#3).}}]}
\renewcommand{\@endtheorem}{\end{trivlist}}

\renewcommand{\@evenfoot}{}
\renewcommand{\@oddfoot}{}

\let\OLDthebibliography\thebibliography
\renewcommand\thebibliography[1]{
  \OLDthebibliography{#1}
  \setlength{\parskip}{0pt}
  \setlength{\itemsep}{0pt plus 0.3ex}
}
\renewcommand*{\@biblabel}[1]{#1.\hfill}

\renewcommand{\@evenhead}%
{%
\begin{otherlanguage}{english}%
\raisebox{0pt}%
[\headheight]%
[0pt]%
{%
\vbox{\hbox to\textwidth{\thepage\strut\hfil\authorseng\hfil}\hrule\vspace{8pt}
}}%
\end{otherlanguage}%
}

\renewcommand{\@oddhead}%
{%
\begin{otherlanguage}{english}%
\raisebox{0pt}%
[\headheight]%
[0pt]%
{%
\vbox{\hbox to\textwidth{\strut\hfil\articleshorttitleeng\hfil\thepage}\hrule\vspace{8pt}
}}\end{otherlanguage}}

\makeatother

\usepackage{fancyhdr}

\fancypagestyle{firstpagestyleeng}{
\fancyhf{}
\headheight=15pt
\fancyhead[CE]{\authorseng \hfill \articleshorttitleeng}
\fancyhead[CO]{\authorseng \hfill \articleshorttitleeng}

}

\fancypagestyle{basestyleeng}{
\fancyhf{}
\headheight=15pt
\fancyhead[LE,RO]{\thepage}
\fancyhead[CE]{\articleshorttitleeng}
\fancyhead[CO]{\authorseng}

}

\newcommand{\paperyear}{2021}
\newcommand{\papervolume}{31}
\newcommand{\papernumber}{1}

\newcommand{\authorseng}{A.\,B.~Veretennikov}

\newcommand{\articletitleeng}{Relevance ranking for proximity full-text search based on additional indexes with multi-component keys}
\newcommand{\articleshorttitleeng}{Relevance ranking for proximity full-text search}

\newcommand{\MSC}{68P20, 68P10}

\newcommand{\paperdoi}{10.35634/vm210110}

\newcommand{\referateng}{The problem of proximity full-text search is considered. If a search query contains high-frequently occurring words, then multi-component key indexes deliver an improvement in the search speed compared with ordinary inverted indexes. It was shown that we can increase the search speed by up to 130 times in cases when queries consist of high-frequently occurring words. In this paper, we investigate how the multi-component key index architecture affects the quality of the search. We consider several well-known methods of relevance ranking, where these methods are of different authors. Using these methods, we perform the search in the ordinary inverted index and then in an index enhanced with multi-component key indexes. The results show that with multi-component key indexes we obtain search results that are very close, in terms of relevance ranking, to the search results that are obtained by means of ordinary inverted indexes.
}

\newcommand{\keywordseng}{full-text search; search engines; relevance ranking; inverted indexes; proximity search; three-component key indexes.}

\newcommand{\datereceive}{11.10.2020}

\newcommand{\contactinformationeng}{Veretennikov Alexander Borisovich,
PhD, Associate Professor, Chair of Calculation Mathematics and Computer Science,
Ural Federal University, pr. Lenina, 51, Yekaterinburg, 620083, Russia. \\
ORCID: \url{https://orcid.org/0000-0002-3399-1889}\\
E-mail: alexander@veretennikov.ru
}

\begin{document}

\titleeng

\noindent
\textbf{Indexing: Web of Science, Scopus.}
\vspace{8pt}

\noindent
\textbf{This is the English translation performed by the author of the original Russian paper.}
\vspace{8pt}

\noindent
\authorseng. \articletitleeng, {\it Vestnik Udmurtskogo Universiteta. Matematika. Mekhanika. Komp'yuternye Nauki}, \paperyear, vol.~\papervolume, issue~\papernumber, \mbox{pp.~132--148.}

\noindent
\href{https://doi.org/10.35634/vm210110}{https://doi.org/10.35634/vm210110}
\vspace{8pt}

\noindent
\href{http://vst.ics.org.ru/journal/article/3051/}{http://vst.ics.org.ru/journal/article/3051/}
\vspace{16pt}

\annotationandkeywordseng

\vspace{1ex}

\makeatletter
\@addtoreset{equation}{section}
\@addtoreset{footnote}{section}
\renewcommand{\section}{\@startsection{section}{1}{0pt}{1.3ex
plus 1ex minus 1ex}{1.3ex plus .1ex}{}}

When full-text search systems are considered, a search query consists of one or several words, and a search result is a list of records. Each record includes information about the document that contains the queried words and information about where in the document the queried words occur. The list of records is sorted, with the most relevant records, i.e., the records that probably most exactly satisfy the needs of the user, are at the beginning of the list. When a proximity full-text search is considered, the documents that contain the queried words near one another are most important. If the queried words occur in a document accidentally in different places, then this document reflects the needs of the user in a minor way, in contrast to a document in which the queried words occur near one another and, therefore, are connected with one another in the context of this document. When a proximity full-text search is implemented, it is required that information about each occurrence of each word be stored in the full-text index \cite{FastKWordProximitySearch,KWordProximitySearchEncrypted}. In other words, the Word-Level index is needed instead of the Document-Level index. Each record in the search results usually contains information about the fragment of text that contains the queried words. Moreover, this fragment of text should have the minimal length among such fragments. This information contains positions of the start of the fragment and of the end of the fragment in the document.

\begin{figure}[tbp]
\setlength{\abovecaptionskip}{1pt}
\setlength{\belowcaptionskip}{-10pt}
\setlength{\abovedisplayskip}{0pt}
\setlength{\belowdisplayskip}{0pt}
\begin{center}
\begin{psfrags}
\small
\includegraphics[width=310pt,height=150pt]{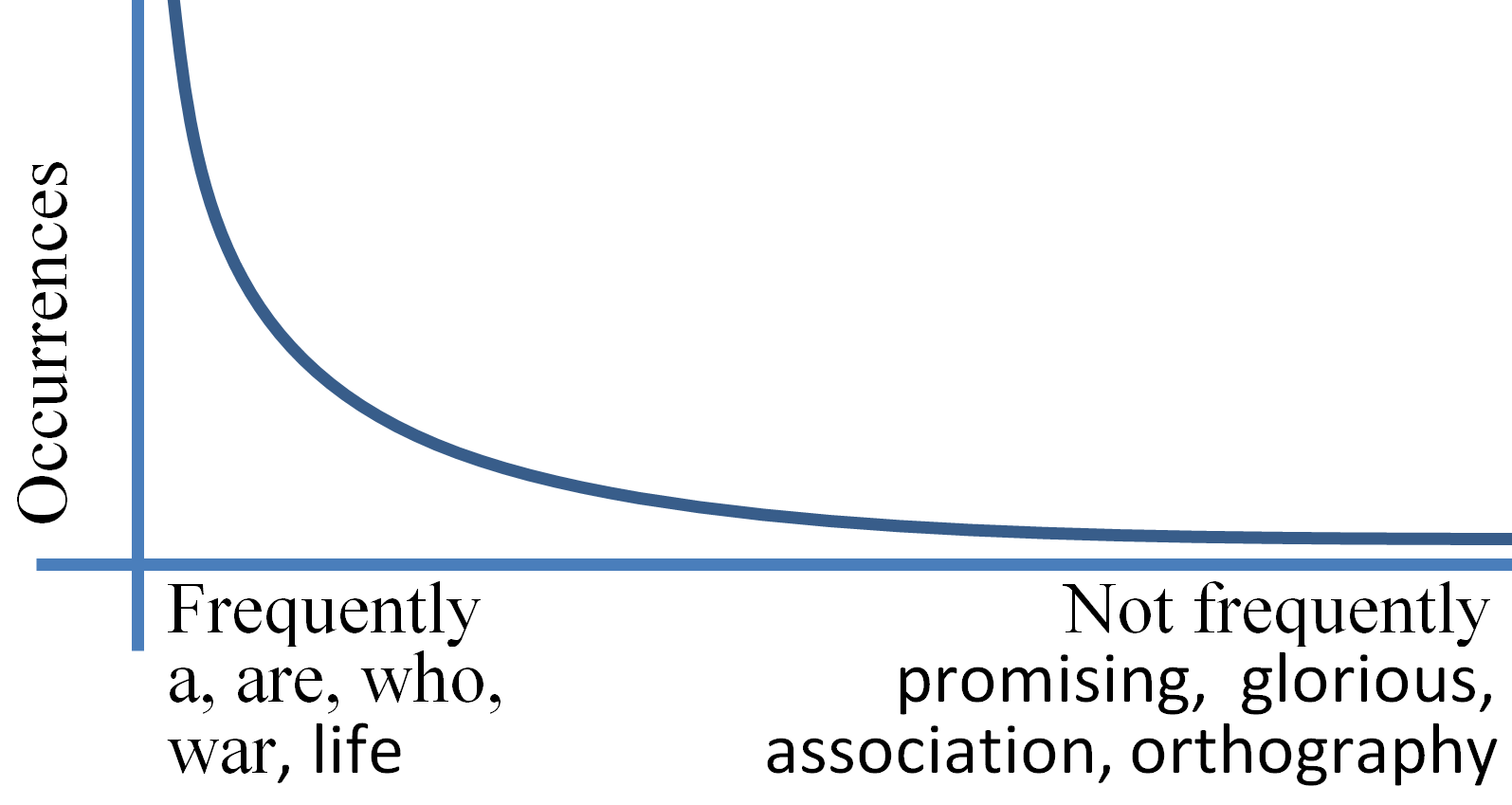}
\end{psfrags}
\end{center}
\caption{An example of word frequency distribution}\label{VeretennikovA-Image-Zipf}
\end{figure}

Words in documents occur with different frequencies. An example of word frequency distribution in texts, an illustration of Zipf’s law \cite{Zipf2}, is presented in Fig.\,\ref{VeretennikovA-Image-Zipf}. Therefore, the search query execution time depends on the total number of occurrences of the queried words in all the indexed documents. We can often see that a search system evaluates queries that consist of only ordinary words quickly, in less than one second. However, if a query contains a frequently occurring word, the search system requires much more time, for example, 20-30 sec., in order to evaluate the query and produce the search results. An example of such behaviour with employment of Lucene is presented in \cite{ProximityFTMultiComponentKeys}. In such cases, the user can say that the system is unstable. 

A search query can be considered a simple inquiry \cite{ResponseTime}. If the query processing time is greater than one second, then the continuity of thought of the user can be interrupted, and, therefore, the performance of the user can be decreased \cite{ResponseTime}. The recommended maximum time of a simple inquiry is two seconds. To resolve this problem, the computational resources can be increased. Another solution is to use additional indexes to increase the search speed. The author of the current work proposed a method of full-text search based on the employment of additional indexes \cite{ProximityFTSG,ProximityFTWithRTG,EffectiveAlgorithThreeKeyCreate}. In the current paper, we show that with these indexes we obtain relevant results. 

The third approach is based on a limitation of acceptable words that the user can search. To achieve this, a list of stop words is defined. Then, these words are excluded from the index and the search. This approach is limited, as the list of stop words cannot be long. 

The fourth approach is early-termination \cite{EarlyTermination,AccessOrdered,10.1145/2808194.2809477,PrecomputedImpacts}. The data in the indexes should be ordered in a specific manner, and specific compatible relevance calculation methods should be applied. If so, then some part of the data can be evaluated as irrelevant and skipped when a search query is executed. However, when a proximity full-text search is required, these methods do not allow us to achieve substantial results \cite{ProximityFTMultiComponentKeys}.

The author of the current paper developed a full-text search method based on the implementation of additional indexes. With these indexes, the search query speed was increased by up to 130 times compared with ordinary indexes \cite{ProximityFTMultiComponentKeys}. Search queries consisted of very commonly occurring words. In \cite{VeretennikovEffectiveSearch}, arbitrary queries were considered. Here we estimated the average number of index records, i.e., postings, that are required to be read from the disk when a search is performed. This number was up to 263 times lower when additional indexes were used, in contrast to the case when only ordinary inverted indexes were employed. The search time can be significantly different when different index parameters or text collections are used. However, these results show that our method based on additional indexes is a perspective method.

The results of the current work are following. When additional indexes are used, the search results are very similar in terms of relevance to the search results obtained with ordinary indexes. This was confirmed by experiments. We used the well-known GOV2 text collection in all experiments. We determined a set of relevance calculation methods, which allowed us to achieve a good level of aforementioned similarity. When these methods are used, the search results obtained with our additional indexes are very similar to the search results obtained with ordinary inverted indexes. We developed a method to compare the different search results obtained with the employment of different indexes. We developed a method to calculate the metrics and statistical values that are required to calculate the relevance when additional indexes are used. We then defined a two-step search method that can be used for text collections, which consist of small documents.

A significant part of the paper is dedicated to the following theoretical questions: a review of different search methods; an examination of the strong sides of the proposed additional indexes; selection of a method that can be used to compare different search results produced from different types of indexes; and a discussion of methods that can increase the search quality when our additional indexes are used. Then the results of the experiments and their analysis are presented.

An inverted index is the data structure that is used for full-text search \cite{InvertedFiles,BlockLinkedListLD}. For each occurrence of each word in each document, a posting is stored in the index. The posting is the record $(ID, P)$. Here, $ID$ is the identifier of the document, and $P$ is the position of the word in the document. An inverted file is the associative array, which allows us to obtain a list of postings for a specific key, for example, for a word. 

When nextword indexes \cite{CombinedIndexes} are used, a key consists of two words. The value of the key is the list of postings, where each posting corresponds to an occurrence of these two words in the text in a consecutive manner, that is, one is immediately after another. A phrase can also be considered as a key \cite{CombinedIndexes}. Such indexes can be used to optimize the phrase search; that is, when the user searches for a document that contains the specified phrase. In this case, the queried words should occur in the document one immediately after another in the correct order. If a document contains the queried words, but another word occurs between the queried words in the text, then this document cannot be found by nextword or phrase indexes. Therefore, the application of these indexes is limited.

Can stop words be excluded from the search? If we exclude some set of high-frequently occurring words from the search, then the performance problem can be solved. However, modern search systems, such as Google, allow the user to search documents that contain any words. Sometimes, high-frequently occurring words can have a specific meaning in the context of a specific search query. In this case, exclusion of such a word from the search can lead to unpredictable effects \cite{ProximityFTMultiComponentKeys, CombinedIndexes}. For example, consider the following search query: ``who are you who''. Here, the second word ``who'' is the name of an English rock band and ``who are you'' is the title of one of their works. In addition, information about stop words can be used for the relevance calculation \cite{AComparisonDAAT}. Our goal is to provide a fast search for any query. Therefore, we include information about all words in our indexes.

When early-termination \cite{EarlyTermination,AccessOrdered,10.1145/2808194.2809477,PrecomputedImpacts} is used, postings are sorted in the index in a specific way. Consider a list of postings. At the beginning of the list, the postings that correspond to the most relevant documents occur, while at the end of the list, the postings that correspond to low relevant documents occur. This approach allows us to stop the reading of the posting list when appropriate criteria are met, for example, when a required number of relevant documents is already found. However, this approach has limited applications to proximity full-text searches. There can always be a document that contains the queried words near one another, but whose postings occur at the end of the posting list of some queried word.

The goal of the current paper is following. When our additional indexes are used, for each word of each document, we store information about other words that occur near this word. A word occurs near another word if the distance between them is less than or equal to $MaxDistance$, where, $MaxDistance$ is a parameter that can have a value of 5, 7, 9 or even greater. However, imagine that some of the queried words occur in a document at a distance that is greater than $MaxDistance$. For example, $MaxDistance$ is equal to 9, and there is a document in which two words that we are searching for occur at a distance of 10 between them. A potential search result will be omitted in the search. This means that when our additional indexes are used, we receive fewer results compared to ordinary indexes. An important question here is the following: are the ``omitted'' results relevant? Most likely, search results in which queried words occur far from one another are not relevant, not important, and can be excluded. In this case, the applicability of our additional indexes will be justified.

In modern relevance calculation methods, this factor has already been taken into account. Let us consider the occurrence of two queried words in a document. In many methods, the relevance score of such an occurrence is inversely proportional to the square of the distance between these words in the document. Such a value is near zero if the words are far from one another. In the following sections, we consider several relevance calculation methods that have been proposed by different authors for proximity full-text search. For each method, we do the following. First, we apply it in the search based on the ordinary inverted index. Then we apply the method in the search in which our additional indexes are used. We consider the search results obtained from the ordinary inverted index as basic. Then, we compare the search results obtained from our additional indexes with the basic search results. We check whether all relevant documents that are presented in the basic results also exist in the results obtained using our additional indexes. We then estimate how these two search result sets are similar to one another in terms of relevance.

\sect{\label{t-relevance-methods}Relevance calculation methods}
\label{resob}

Let us consider some search results, which are sorted in terms of relevance. The documents occur in the search results in the order of relevance, from a high relevance to a low relevance. This can easily be done if a relevance function exists. Such a function that produces a number, that is, the relevance score, for every search result, can be used. For example, BM25 \cite{BM25} or TF-IDF \cite{TFIDF}.

When the proximity full-text search is considered, it is often assumed that the relevance score of the document is inversely proportional to the square of the distance between the queried words within the document \cite{ProximityWithPair}. Let us consider the search query that consists of two words, letting $A$ and $B$ be the positions of these words in the document. Then, 
$TP = 1 / (A - B)^2,$
where $TP$ is the relevance score of the document in the context of the proximity between the queried words in the text. This approach can easily be extended for a search query $Q$, which consists of $n$ words \cite{ProximityFTWithRTG}. Let $X=X_1, \dots, X_n$ be positions of the queried words in the document. Then, 
$$TP(X)= 1 \Big/ \Big(|A(X)-B(X)| - (n-2) \Big)^2,
\text{ in which } A(X)=\min\limits_{1 \leqslant i \leqslant n} X_i, \: B(X)=\max\limits_{1 \leqslant i \leqslant n} X_i.$$

The relevance of a document can be defined in the following way. First, the search results should be sorted in accordance with the value of $TP$. Then, each sub-list of the search results, where each search result has the same value of $TP$, should be sorted in accordance with the value of BM25 or TF-IDF. These methods are denoted as $R_{TP,\text{BM25}}$ and $R_{TP,\text{TF-IDF}}$, respectively.

On the other hand, when the proximity full-text search is considered, the relevance function can be specified as the weighted sum of several other functions. In \cite{ProximityWithPair}, the following relevance function was proposed:
$R = \alpha \cdot SR + \beta \cdot IR + \gamma \cdot TP$,
in which 
$\alpha, \beta,  \gamma \geq 0, \alpha + \beta + \gamma = 1.$

Here, $SR$ is a static document rank, such as PageRank \cite{PageRank}, $IR$ is an information retrieval rank, such as BM25 or TF-IDF, in which the queried words are taken into account, and $TP$ is a proximity rank, in which proximity information, that is, information about distances between the queried words in the document, is taken into account. The parameters, $\alpha, \beta$, and $\gamma$ are the weights of the $SR$, $IR$, and $TP$, respectively. The values of $SR$, $IR$, and $TP$ are normalized, that is, each value belongs to the $[0, 1]$ interval. We denote this method as $R_{WeiSum,\alpha, \beta, \gamma}$, or, $R_{WeiSum, 0.1, 0.4, 0.5}$, for example, when a specific values of the parameters are selected. When the following experiments were conducted, we used a text collection without the calculated $SR$ values. Therefore, we used $\alpha=0$. We used BM25 as the $IR$ rank function.

Some ranking functions are defined for the entire document, for example, BM25. Other functions are defined in the context of a fragment of a text within which the queried words occur, and such functions can depend on the distance or the order of the queried words in the text. Therefore, more than one query result can exist for one document. We can have several search results for a single document, with different values of the relevance score. For example, if the queried words occur in the document in two different places, we have two search results. For these two search results, we have the same value of BM25 but can have different values of $TP$. In the search result list, we can have several search results for the same document. For these search results, we can have the same or different values for the relevance score, which depends on the selected relevance function.

The relevance score can be defined based on intervals. In one of the methods \cite{HigherOrderProximityModeling}, all variants of the queried word occurrences should be considered, and then the relevance score of the document should be calculated. The method in \cite{HigherOrderProximityModeling} is based on the approaches \cite{ClarkeIntervals,TermProximityPerspective}. Let $X=X_1, \dots, X_n$ be some positions of the queried words in the document. Then $[A(X), B(X)]$ is the interval that covers the queried words. The interval should be minimal; that is, it should contain all of the queried words, but should not contain an interval of a lesser length that also contains all of the queried words. The document can contain several suitable intervals that can be far from one another. For each interval a relevance score should be calculated. Then the relevance score of the document is defined as the sum of the calculated relevance scores of these intervals. The final relevance score is defined for the entire document, not just for a part of it. In general, intervals that contain some subset of the queried words can also be considered. Different methods of calculating the interval relevance score can be used.

Let us define the following subsets of documents \cite{ClarkeIntervals}: $D_1, D_2, \dots, D_{|Q|}$. Here, $|Q|$ is the length of query $Q$. Let $D_1$ be a subset of documents, each of which contains one of the queried words, let $D_2$ be a subset of documents, each of which contains two of the queried words, etc. Let $D_{|Q|}$ be a subset of documents, each of which contains all of the queried words. When a list of relevant documents is determined, these subsets are examined in reverse order, starting with $D_{|Q|}$. If the required number of relevant documents is found in $D_{|Q|}$, then the remaining subsets are no longer needed. That is, first, the documents that contain all of the queried words are important. The relevance score or weight of the interval $I=[p, q]$ in document $D$, where $p$ is the start of $I$ and $q$ is the end of $I$, are defined as follows:
$Score_{\text{Clarke et al.}}(I,D) = min\left( K / (q-p+1), 1 \right)$, where $K=16$.

The relevance score of the document is defined as follows. Let us search for all suitable intervals in the document. Then we calculate the relevance score for each interval and sum these values. Please note that the relevance score of an interval here is inversely proportional to the length of the interval (not to the square of the length of the interval).

We consider such relevance calculation functions in a limited way. We consider only intervals that contain all of the words of the search query. To consider all possible intervals, we need to implement a complex search algorithm. We already implemented several different search algorithms that apply to different query types (see later). Because, for different query types we should use different additional indexes \cite{ProximityFTWithRTG, ProximityFTMultiComponentKeys}. In \cite{ClarkeIntervals}, first, the documents that contain all of the words of the search query are processed; therefore, the aforementioned limitation is not critical for an analysis of the method from \cite{ClarkeIntervals}. We denote this method as $R_{IntervalSum}$. We then assume the following. If the applicability of our additional indexes will be shown in these limited conditions, then they will work fine and without these limitations. In the future, we plan to implement a more complex search algorithm. In the future algorithm, we plan to take all intervals that contain some of the words of the search query into account.

In \cite{TermProximityPerspective}, the text of the document is scanned from left to right. In this process, a list of intervals is obtained. If some words of the search query are near one another in the text, then these words are placed in one interval. Different intervals contain different numbers of queried words. Let us consider the search query $Q$ and document $D$.  Then, 
\begin{displaymath}
\setlength{\abovedisplayskip}{2pt}
\setlength{\belowdisplayskip}{2pt}
\text{BM25}(Q, D) = \sum\limits_{e \in Q} W_e \frac{TF(D,e)(1+k_1)}{TF(D,e)+K}, \text{ where }
K=k_1 \times \left(1-b+\frac{b \times |D|}{avg(|D|)}\right),
\end{displaymath}

where $k_1$ and $b$ are parameters and $TF(D,e)$ is the term frequency of the lemma or word $e$ in document $D$. Usually, $TF(D,e)$ is the total number of occurrences of $e$ in $D$. Let $|D|$ be the length of $D$, that is, the total number of words in the document, and let $avg(|D|)$ be the average document length in the indexed collection of the documents. For the value of $W_e$, IDF can be used, as can other metrics \cite{HigherOrderProximityModeling}. 

In \cite{TermProximityPerspective}, a similar formula was used, in which $TF(D,e)$ was replaced by $rc(D,e)$. The latter $rc$ is defined by the sum of the relevance scores of intervals containing the word or lemma $e$. The relevance score of an interval $I$ is defined by the following formula:
$$Score_{\text{Song et al.}}(I,D) = \frac{n_i^{\lambda}}{(q-p+1)^{\gamma}}.$$

This is done in similar way in \cite{ClarkeIntervals}. Here, $n_i$ is the number of queried words that occur in the interval $I$, and $\lambda$ and $\gamma$ are the parameters.

In \cite{HigherOrderProximityModeling}, when the relevance score of an interval is calculated, the IDF weights of the words, which occur at the start or at the end of the interval (that is, the interval’s boundary terms), are taken into account and BM25-like constructions are used. The authors of \cite{HigherOrderProximityModeling} stated that BM25 generally works well accordingly \cite{ImprovementsBM25}. The relevance score of the interval $I = [p,q]$ is defined as follows:
$$Score_{\text{Lu et al.}}(I,D) = W_l \cdot W_r  \cdot (q - p + 1)^{-2},$$
where $W_l$, and $W_r$ are the IDF weights of the lemma that occurs at the start of the interval and the lemma that occurs at the end of the interval, respectively. This leads to the following: the relevance scores of the intervals, in which the boundary terms are frequently occurring words, decrease. The relevance score of the document is then calculated by the BM25-like formula, as in \cite{TermProximityPerspective}. Let $\mathcal{I} = \mathcal{I}(Q')$ be a set of intervals, where each interval contains a subquery $Q'$. Then, 
\begin{displaymath}
\setlength{\abovedisplayskip}{2pt}
\setlength{\belowdisplayskip}{2pt}
Score_{\text{Lu et al.}}(Q',D) = 
\frac{\sum_{I \in \mathcal{I}} Score_{\text{Lu et al.}}(I,D) \cdot (1 + k_1)}{\sum_{I \in \mathcal{I}} Score_{\text{Lu et al.}}(I,D) + K'}, \text{ where }
K'=K \cdot \left( \sum_{e \in Q'} min(W_e,1) \right),
\end{displaymath}
and $W_e$ is the IDF weight of $e$. Let $\lambda$ be a parameter with a representative example value of 0.4 as in \cite{HigherOrderProximityModeling}. To calculate the relevance score $Rel$ of the document, all subqueries of the search query $Q$ are considered. Two sets of subqueries are also considered. Let $\mathcal{Q}$ be the set of all subqueries of the search query $Q$, and let $\mathcal{Q}'$ be the set of all subqueries of the search query $Q$, in which the order of the words is the same as in $Q$. Then,
\begin{displaymath}
\setlength{\abovedisplayskip}{2pt}
\setlength{\belowdisplayskip}{2pt}
Rel_{\text{Lu et al.}}(Q,D) = (1 - \lambda) \cdot BM25(Q,D) +  
\end{displaymath}
\begin{displaymath}
\setlength{\abovedisplayskip}{2pt}
\setlength{\belowdisplayskip}{2pt}
\lambda \cdot \sum\limits_{Q' \in \mathcal{Q}} Score_{\text{Lu et al.}}(Q',D) +
\lambda \cdot \sum\limits_{Q' \in \mathcal{Q'}} Score_{\text{Lu et al.}}(Q',D).
\end{displaymath}

We calculate the approximate value of this function, only considering intervals that contain all of the words of the search query. We denote this method as $R_{IntervalOpt}$.

In modern search systems, a two-level relevance calculation process is often used \cite{TwoStageL2R, LambdaMART}. Two ranking mechanisms are applied one after another. The first mechanism forms a preliminary result set and excludes obviously non-relevant results. Already considered relevance functions, such as BM25, can be used here. The second mechanism processes the preliminary result set and calculates the final relevance scores of the documents based on different features of each search result, such as: BM25, TF-IDF, document length, PageRank, the first position of a queried word in the document, distance between the queried words in the document, etc. For every document or search result, we have a vector of numbers that represents these features of the search result. 

The second mechanism can be implemented using machine learning techniques, such as artificial neural networks \cite{NNL2R}, genetic algorithms \cite{GeneticL2R}, and gradient boosting \cite{LambdaMART}. To construct the relevance function, a test document collection is used with a specific set of test queries. For each test query, the list of relevant documents is given, and for each of these documents, the feature vector and the relevance score values are also presented. After the relevance function, such as an artificial neural network, for example, is constructed, this function allows us to calculate the final relevance score of arbitrary documents in the context of an arbitrary search query based on the feature vector of the document. The first mechanism is usually low-cost and works quickly. While the second mechanism works slowly, it improves the quality of the search results.

In the current work, we analyze the relevance functions, which can be used as part of the first mechanism. Their results can be used for a second-level mechanism. A second-level mechanism is based on the results of a first-level mechanism. Therefore, if a first-level mechanism provides similar results in both cases, that is, when additional indexes are used and when ordinary indexes are used, then the second-level mechanism will also work fine in both cases. Let us consider several relevance functions, and for each function, conduct an experiment as follows.

Let us select a test query set. For each query we perform the following.

\vspace{8pt}
\begin{itemize}
\item	We evaluate the query using the ordinary inverted index. The first result set is obtained.
\item	We sort the first result set according to the relevance function.
\item	We evaluate the query using our additional indexes. The second result set is obtained.
\item	We sort the second result set according to the relevance function.
\item	We compare both result sets.
\end{itemize}

If both results sets are similar, then our additional indexes can be applied for a full-text search with this relevance function, where the search is performed without a loss of quality. 

We consider the following relevance functions.

\vspace{8pt}
\begin{itemize}
\item	$R_{TP,\text{BM25}}$, $R_{TP,\text{TF-IDF}}$, 
\item	$R_{WeiSum, 0, 0.75, 0.25}$, $R_{WeiSum, 0, 0.5, 0.5}$, $R_{WeiSum, 0, 0.25, 0.75}$, $R_{WeiSum, 0, 0.1, 0.9}$,
\item	$R_{IntervalSum}$ -- sum of the relevance scores of the intervals,
\item	$R_{IntervalOpt}$ -- BM25-like constructions and sum of the relevance scores of the intervals,
\item	$R_{IntervalSumSq}$ -- sum of the relevance scores of the intervals, such as $R_{IntervalSum}$, but the relevance score of an interval is defined as for $R_{WeiSum, 0, 0.75, 0.25}$. That is, the weight of an interval is inversely proportional to the square of the length of the interval, but not to the length of the interval.
\end{itemize}

These functions represent the following relevance calculation methods.

\vspace{8pt}
\begin{enumerate}
\item[1] First, define the list of relevant documents using the proximity between the queried words in the text as the criteria of relevance. Second, refine the list of relevant documents using a relevance function that does not depend on the positions of the queried words in the text ($R_{TP,\text{BM25}}$, $R_{TP,\text{TF-IDF}}$).
\item[2] Use a weighted sum in which the following components are presented. First, a component that takes the proximity between the queried words in the text into account. Second, a component that does not depend on the positions of the queried words in the text ($R_{WeiSum, 0, 0.75, 0.25}$, $R_{WeiSum, 0, 0.5, 0.5}$, $R_{WeiSum, 0, 0.25, 0.75}$ and $R_{WeiSum, 0, 0.1, 0.9}$).
\item[3] Let us consider a document. The relevance score is defined for entire document. Intervals that contain the queried words are then collected. The lengths of these intervals are used to obtain the relevance score of the document ($R_{IntervalSum}$ and $R_{IntervalSumSq}$).
\item[4] Consider a set of intervals, each of which contains the queried words. Information about the lengths of these intervals is then combined with IDF information in a BM25-like manner ($R_{IntervalOpt}$).
\end{enumerate}

\sect{\label{t-search-result-compare}Search result comparison}

Let us consider a search query $Q$. Let $Ideal$ be the list of search results obtained in the search in the ordinary inverted index, and let $Instance$ be the list of search results obtained in the search when our additional indexes are used. Every item in a list of search results contains the following fields.

$ID$ -- the identifier of a document.

$P$ -- the position of the start of a fragment of the text that contains the queried words.

$L$ -- the length of the aforementioned fragment of the text.

$R$ -- the relevance score of the search result (a floating-point number).

We need to compare these two lists of search results as two vectors. Moreover, we need to obtain a numerical value that represents the degree to which these two lists are different from one another. We use the following metrics for this.

\vspace{4pt}
Levenshtein distance \cite{Lev65}.

\vspace{4pt}
$Precision$ \cite{PrecisionRecall}.

\vspace{4pt}
$NDCG$ (Normalized Discounted Cumulative Gain) \cite{Jarvelin:2,LETOR}.

\vspace{4pt}
To use these metrics, we need to define conditions when two records $X$ and $Y$, both of which have structures $(ID,P,L)$, are equal to one another. These conditions are the following.

1) $X.ID = Y.ID$, 

2) $EP(X) = EP(Y)$.

Here $EP(V) = V.P$ when $V.L < LRD$, and $EP(V)=-1$ otherwise.

$LRD$ is a parameter with a representative example value of 50.

We use condition 2) to address the following issue. If two words occur in a text at a large distance ($>50$, for example), then the exact value of this distance is irrelevant. It only matters that these two words are far from one another. Let us consider two search results $X$ and $Y$, which are related to the same document, that is, $X.ID = Y.ID$. If in both cases, that is, $X$ and $Y$, the queried words occur in the document far from one another, then we consider that $X$ and $Y$ are equal. The distance between the queried words in the document can be different for $X$ and $Y$, but this does not matter. Please note that modern search systems present to the user a fragment of text that contains the queried words for every search result. For Google, the length of such a fragment is approximately 15-30 words. 

The aforementioned definition is important when a two-step search process (see more about this later) is employed when we search with additional indexes \cite{ProximityFTSG}. 

Let us consider a number $N$. Let $Ideal_N$ be the list that consists of the first $N$ elements of $Ideal$. Similarly, let $Instance_N$ be the list that consists of the first $N$ elements of $Instance$. Let us compare $Ideal_N$ and $Instance_N$. For a specific value of $N$, we denote $Precision$ as $P@N$.

$$Precision(N) = P@N = |Ideal_N \cap Instance_N| / |Instance_N|.$$

We consider the list $Ideal_N$ as the list of relevant documents. This list is obtained when the search is performed with an ordinary index.

We estimate the relevance of the list $Instance_N$ in relation to the $Ideal_N$.

We then need to calculate $P@N$. We calculate the number of relevant documents that are contained in $Instance_N$. That is, the number of $Instance_N$ elements that are also contained in $Ideal_N$. We should divide this number on the length of $Instance_N$.

We also calculate the Levenshtein distance between $Ideal_N$ and $Instance_N$.

For a specific value of $N$, we denote the value of $DCG$ (Discounted Cumulative Gain) as $DCG@N$.

\begin{displaymath}
\setlength{\abovedisplayskip}{2pt}
\setlength{\belowdisplayskip}{2pt}
DCG@N = \sum\limits_{i=1}^N \frac{2^{Rel(Instance[i])}-1}{log_2 (i+1)}, 
\end{displaymath}

where $Rel(Instance[i])$ is the relevance score of the $i$-th record of $Instance$. The numeration of $Instance$'s elements begins with 1. The value of $Rel(x)$ is defined based on the relevance scores of the elements of $Ideal$. However, $x$ can be an element of an arbitrary search result list. Let $x$ be a record $(ID,P,L,R)$, an element of $Instance$. We need to define $Rel(x)$. We need to find a record $y$ in $Ideal$ that is equal to $x$. To compare $x$ and $y$, we use the values of the $ID, P, L$ fields. If such a record $y$ is found in $Ideal$, then $Rel(x) = y.R$.

For a specific value of $N$, we denote the value of $IDCG$ (Ideal Discounted Cumulative Gain) as $IDCG@N$. This value is then calculated in the same way as $DCG$ but for $Ideal$.

\begin{displaymath}
\setlength{\abovedisplayskip}{2pt}
\setlength{\belowdisplayskip}{2pt}
IDCG@N = \sum\limits_{i=1}^N \frac{2^{Rel(Ideal[i])}-1}{log_2 (i+1)}, 
\end{displaymath}

where $Rel(Ideal[i])$ is the relevance score of the $i$-th record of $Ideal$. The numeration of $Ideal$'s elements also begins with 1. That means that $Rel(Ideal[i]) = Ideal[i].R$.

For a specific value of $N$, we denote the value of $NDCG$ (Normalized Discounted Cumulative Gain) as $NDCG@N$ and calculate it as follows.
$$NDCG@N=(DCG@N) / (IDCG@N).$$

Please note that $0 \leq NDCG@N \leq 1$.

If the value of $NDCG@N$ is near 1, then our goal is achieved. This means that when we search with our additional indexes, our search results are relevant and similar to the search results that are obtained with the ordinary index. The notion of similarity between different search results is defined here in terms of relevance.

Let us consider $R_{TP,\text{BM25}}$ and $R_{TP,\text{TF-IDF}}$. In these cases, the relevance scores, that is, the values of the $R$ component, are not defined in the search results. This is because the two-level sorting process is implemented here. Therefore, in these two cases we define $Ideal[i].R = 1/i$.

$NDCG$ is better than $Precision$ and the Levenshtein distance, because to calculate $NDCG$, we take the numerical values of the relevance scores into account. Let us consider the following example. Let the first two documents in $Ideal$ be highly relevant. Let us imagine that $Ideal$ also contains $100$ low relevance documents. The relevance scores for these low relevance documents are near 0. Let $N$ be 30, and let us calculate $NDCG$. The absence of some low relevance documents in $Instance$ will not be a problem.

However, when we calculate $Precision$ or the Levenshtein distance the following occurs. The absence of some low relevance documents in $Instance$ can lead to an incorrect interpretation of these metrics. These metrics can show to us that we have a lower search quality than we truly have. When $NDCG$ is calculated, we do not have such issue.

Let us examine a dynamic of changes in the $Precision$ and the Levenshtein distance when different values of $N$ are considered. Let $N$ be a small number. In this case, we see to what degree the results of the search with the additional indexes contain the same highly relevant documents obtained by a search with the ordinary index. In other words, a high value of $NDCG@N$ when $N$ is an arbitrary number should correspond to high values of $Precision$ and the Levenshtein distance that are calculated when $N$ is a small number. However, the values of $Precision$ and the Levenshtein distance can decrease with an increase in $N$, but that does not mean that the search quality is low.

The $NDCG$ metric is considered a primary metric and is often used when search quality is investigated, when some search result lists are compared with the ideal search results. The ideal search results can often be selected manually. However, in our case we consider the search results obtained by the search in the ordinary index to be ideal search results. In the following experiments, we evaluate some number of test search queries and calculate the average values of the aforementioned metrics.

\sect{\label{t-lemmatization}Lemmatization and additional indexes}

We employ a morphological analyzer. For a word, the analyzer provides a list of basic forms, that is, lemmas. The dictionary of the analyzer contains approximately 92 thousand English lemmas. Let $FL$ be the list of all lemmas. Let $FL$ be sorted in a decreasing order of lemma occurrence frequency. Let $FL$-number of the lemma $x$, that is, $FL(x)$, be the ordinal number of $x$ in $FL$.

Let us consider two arbitrary lemmas, $x$ and $y$. We define that $x < y$ when $FL(x) < FL(y)$.

Let the first $SWCount$ elements of $FL$ be stop lemmas. For example, ``war'', ``time''.

Let the next $FUCount$ elements of $FL$ be frequently used lemmas. For example, ``beautiful'', ``red''.

Let all the following lemmas of $FL$ be ordinary lemmas. For example, ``glorious'', ``promising''.

Here, $SWCount$ and $FUCount$ are parameters.

GOV2 \cite{TREC:2006} text collection was used for the experiments. This text collection contains mostly English documents. We only use the English dictionary, $SWCount = 500$ and $FUCount = 1050$. 

The value of $SWCount = 500$ is near $421$, which was used in \cite{Fox:1989:SLG:378881.378888}. Please note that a word can have several lemmas. For example, the word ``mine'' has two lemmas, namely, ``mine'' and ``my''.

We employ several types of additional indexes \cite{ProximityFTWithRTG}.

The three-component key $(f,s,t)$ index is the list of the occurrences of lemma $f$ in which lemmas $s$ and $t$ both occur in the text at distances that are less than or equal to the $MaxDistance$ from $f$. Here, $f$, $s$ and $t$ are stop lemmas and $f \leq s \leq t$. Every posting record in the index has the format $(ID,P,D1,D2)$. Here, $ID$ is the identifier of a document, and $P$ is the position of lemma $f$ in the document, for example, the ordinal number of the word in the document, $D1$ is the distance between $f$ and $s$ in the document, and $D2$ is the distance between $f$ and $t$ in the document.

The two-component key $(w,v)$ index is the list of occurrences of lemma $w$ for which lemma $v$ occurs in the text at a distance that is less than or equal to the $MaxDistance$ from $w$. Here, $w$ is a frequently used lemma and $v$ is a frequently used or ordinary lemma. Every posting record in the index has the format $(ID,P,D)$. Here, $ID$ is the identifier of a document, $P$ is the position of lemma $w$ in the document and $D$ is the distance between $w$ and $v$ in the document.

The third additional index is the ordinary index with $NSW$ (near stop word) records. This index contains posting lists for frequently used and ordinary lemmas. For every lemma $x$, the list of records $(ID, P, NSW)$ is stored in the index. Here, $ID$ is the identifier of a document, and $P$ is the position of lemma $x$ in the document. $NSW$ is the $NSW$ record, which contains information about all the stop lemmas that occur in the text near position $P$, that is, at a distance that is less than or equal to $MaxDistance$, from $P$.

For example, let us consider the $(ID, P, NSW((war,3),(time,-2)))$ record. In the document, the stop lemma $time$ occurs at a distance of $(-2)$ from $P$, and the stop lemma $war$ occurs at a distance of $3$ from $P$.

The key points of our research are the following. 

\vspace{4pt}
We store information about all words, including those that are frequently occurring, in the indexes. 

\vspace{4pt}
We employ Word-Level indexes, and we use easy updatable indexes. 

\vspace{4pt}
For each key, several data streams can be used. 

\vspace{4pt}
In the search, each posting list, for each key, is read entirely. 

\vspace{4pt}
We use the DAAT (Document-at-a-time) approach \cite{AComparisonDAAT}.
\vspace{4pt}

We have described the procedure of search query processing in \cite{ProximityFTWithRTG} and \cite{VeretennikovEffectiveSearch}. This procedure should be improved now because we need to calculate the relevance. To do so, the following values are needed.

Let $D$ be a document and $x$ be a lemma. Let $TF(D,x)$ be the number of occurrences of $x$ in $D$, and let $DF(x)$ be the number of documents that contains $x$.

We can calculate $TF(D,x)$ as follows.

If we read the posting list of $x$ from the ordinary index with $NSW$ records, we can then calculate $TF(D,x)$ in the process of this reading for each document. The value of $DF(x)$ will be calculated after the posting list is read entirely. Please note that $NSW$ records are stored in a separate data stream. They can therefore be skipped when they are not needed for a specific search query.

Let us consider stop lemmas and all other lemmas $x$ that satisfy the condition \mbox{$FL(x) < TS$}. For these lemmas we use an additional $DTA$ table. For each such lemma $x$, we perform the following. We store $TF(D, x)$ for each document $D$ and $x$ in $DTA$. We also store $DF(x)$ in $DTA$. Here, $TS$ is a parameter, $TS \geq SWCount + FUCount$. We build $DTA$ in the index construction process, and we store $DTA$ in operational memory when searches are performed. The amount of available operational memory is the primary factor that affects the value of $TS$.

When an $(f, s, t)$ or $(w, v)$ posting list is read, we cannot calculate $TF(D, x)$ or $DF(x)$ for any lemma. This is because $(f, s, t)$ indexes contain information about only such occurrences when $f$, $s$ and $t$ were near one another in the text. Similarly, $(w, v)$ indexes contain information about only such occurrences when $w$ and $v$ were near one another in the text. We have the same situation when we process $NSW$ records and reconstruct postings for stop lemmas from them. In all these cases we use $DTA$. There is also another table, namely, $DL$, which is also stored in operational memory. For each document $D$, the value of $DL(D)$ is the length of $D$ in words. Let $AvgDL$ be the average document length in words, and let $DC$ be the total number of documents.

For each search algorithm, the values of $TF(D, x)$ and $DF(x)$ should be available for every queried lemma. These values are used to calculate BM25 and TF-IDF.

Let us consider the following query types.

QT1. The query contains only stop lemmas. We use $DTA$ to obtain the values of $TF(D, x)$ and $DF(x)$. We employ $(f, s, t)$ indexes.

QT2. The query contains only frequently used lemmas. We use $DTA$ to obtain the values of $TF(D, x)$ and $DF(x)$. We employ $(w, v)$ indexes.

QT3. The query contains only ordinary lemmas. We obtain the values of $TF(D, x)$ and $DF(x)$ when we read the posting lists. We use the ordinary index and skip the $NSW$ records.

QT4. The query contains one or several stop lemmas. The query also contains frequently used and/or ordinary lemmas. Let $w$ be some queried lemma with the following conditions. The lemma $w$ should be a frequently used or ordinary lemma. Lemma $w$ should have minimal occurrence frequency among the queried lemmas. For lemma $w$, we use the ordinary index and read the $NSW$ records. Information stored in the $NSW$ records allows us to take all stop lemmas, which are presented in the search query, into account.

Let us consider another non-stop queried lemma $v$. The word index $i$ of lemma $v$ in the query should be different from the word index of the first occurrence of lemma $w$ in the query. We can consider one of the following cases.

The notion word index here means the following. Consider the search query ``who are you who''. The search query consists of four words. Every word has its index in the query; for example, the first word is ``who'', the second is ``are'', the third is ``you'' and the fourth is ``who''. To evaluate the query, all words of the query and their lemmas should be processed.

1) If $v$ is a frequently used lemma, then we can perform the following. We use the $(w, v)$ index instead of the $(v)$ index. Thus, the $(w, v)$ index allows us to obtain the list of occurrences of lemma $v$, for which lemma $v$ occurs near lemma $w$ in the text. We use $DTA$ to obtain the values of $TF(D, v)$ and $DF(v)$.

2) If $v$ is an ordinary lemma, $FL(v) < TS$ and the search query contains a frequently used lemma $x$ with the word index $j \neq i$, then we can perform the following. We use the $(x, v)$ index instead of the $(v)$ index, and we use $DTA$ to obtain the values $TF(D, v)$ and $DF(v)$.

3) If $v$ is an ordinary lemma, $FL(v) \geq TS$, then we can use the ordinary index $(v)$. However, the $NSW$ records should be skipped. We obtain the values $TF(D, v)$ and $DF(v)$ when we read the $(v)$ posting list.

QT5. The search query contains at least one frequently used lemma, namely, $w$ and several ordinary lemmas. This case is similar to QT4. We can proceed here as in QT4. However, $NSW$ records should be skipped entirely. 

We also have an alternative algorithm here. The Main-Cell algorithm \cite{VeretennikovEffectiveSearch} can be implemented as follows. Let $w$ be a queried lemma with the following conditions. Lemma $w$ should be a frequently used lemma, and should have a minimal occurrence frequency among the frequently used queried lemmas. Let us consider all other queried lemmas, and let $v$ be some other queried lemma. We can use the $(w, v)$ index. This $(w, v)$ index contains the list of occurrences of lemma $w$, for which lemma $v$ occurs near lemma $w$ in the text. Therefore, we have a set of indexes. For each queried lemma, except $w$, we begin to read the index $(w, v)$. The data in the index are sorted in an increasing order. We shift in each index in such a way that the current posting in each index has the same value, that is, $(ID, P)$. If we can do that, then we have a place in the document $ID$ where all required lemmas are presented near one another. We have the search result and store it in the search result list. We obtain the values $TF(D, v)$ and $DF(v)$ in one of the following ways. If $FL(v) < TS$, then we use $DTA$. Otherwise, we should read the $(v)$ posting list from the ordinary index, skipping the $NSW$ records. The latter posting list is not used for the search. We only use it to produce the values of $TF(D, v)$ and $DF(v)$. This is the third approach to obtain the values of $TF(D, v)$ and $DF(v)$ for a specific lemma $v$ when required.

For example, let us consider a search query ``Scalable Vector Graphics''. 

Here, \mbox{$FL(Scalable)=-1$}, $FL(Vector)=3227$, and $FL(Graphics)=1075$, ``Graphics'' is a frequently used lemma, and the other lemmas are ordinary. We use the (graphics, scalable), (graphics, vector) indexes for the search. In addition, the (scalable) index is used to obtain the values of $TF(D,v)$ and $DF(v)$ for $v = scalable$. The $NSW$ records are skipped.

Therefore, a larger value of $TS$, for example, $5000$, makes the search faster.

\sect{\label{t-two-phase-search}Two-step search}

If the queried words occur in a document at a distance that is greater than $MaxDistance$, then our additional indexes cannot help find this document. To solve this problem, we can perform two searches \cite{ProximityFTSG}. The first search is the proximity search. The second search is the non-proximity search; that is, we search documents that contain the queried words anywhere in the text. For the first search, we use our additional indexes, which are Word-Level indexes. For the second search, we only use the Document-Level index. Let us consider a document. In the Document-Level index, for every word in the document, we only store information about the first occurrence of the word. This approach is successful when documents are relatively large, for example, approximately 300 KB or greater. In this case, the non-proximity search works fast. Such metrics as BM25 \cite{BM25} and TF-IDF \cite{TFIDF} are calculated for the entire document. Therefore, all documents that are relevant by these metrics will be found in the second search. The first search ensures that all documents that are relevant in the proximity sense, will also be found.

Let us consider the ordinary index with $NSW$ records. For each key, we store the list of postings, where every posting has the structure $(ID, P, NSW)$. This list of postings can be stored in three data streams, namely, $(ID)$, $(P)$ and $(NSW)$. This allows us to only read the necessary information from the index. If we need a non-proximity search, then we only read the $(ID)$ data streams.

Let us consider a queried lemma $v$. If $FL(v) \geq TS$, then $TF(D,v)$ and $DF(v)$ are obtained in the first search and cached for the second search. If $FL(v) < TS$, then $DTA$ is used. If the posting list is short, then we can only create two data streams in the index, namely, $(ID, P)$ and $(NSW)$. In this case, the total number of disk operations that are required for the index construction can be decreased.

In the ordinary index with $NSW$ records, the keys are frequently used and ordinary lemmas. Let us also consider stop lemmas as keys for this index. However, for stop lemmas, the $NSW$ records are not stored. Moreover, for a stop lemma, only information about the one occurrence, namely, the first occurrence, is stored per document. For a text collection that consist of large documents, the two-step search process is organized in the described manner.

However, GOV2 text collection consists of small documents. In GOV2, the average text length of documents is approximately $7$ KB. In this case, the non-proximity search can be slow. Let us consider a frequently occurring word, and let us search this word. When a large document is considered and we switch from the Word-Level index to the Document-Level index, the search speed increases significantly. This is because, in order to process one document, we need to read only one posting record instead of a long list of posting records. For smaller documents, the search speed is increased, to a lesser degree.

To make the second search faster, we implement the following optimization. Let us consider a search query $Q$. Let $v$ be a lemma of $Q$ where the condition $FL(v) < TS$ is met. Then we can evaluate $Q$ excluding $v$ and obtain the search results. Then, we post-process the search results. We should exclude any document that does not contain $v$ from the search results. If $TF(D,v)=0$, then the document $D$ does not contain the lemma $v$.
 
Therefore, all lemmas $v$ that satisfy the condition $FL(v) < TS$ can be excluded from $Q$. However, the new version of $Q$ should contain at least one lemma. If for each queried lemma $v$ the condition $FL(v) < TS$ is satisfied, then we should select one lemma, which remains in $Q$. This remaining lemma is a queried lemma with a maximum $FL$-number among the queried lemmas, that is, a lemma, which have a lesser occurrence frequency in the texts.

Let us consider the search query ``Scalable Vector Graphics''. 

In this query, $FL(Scalable) = -1$, $FL(Vector) = 3227$, and $FL(Graphics) = 1075$. We use $TS=5000$. Therefore, the new query should be ``Scalable''. We evaluate this query and obtain the search results. Then, we filter the search results. We exclude all documents that do not contain ``Vector'' and ``Graphics'' from the search results.

When this optimization is used, the second search becomes significantly faster. Therefore, the two-step search process can now also be applied to collections that consist of small documents, for example, to GOV2.

Let us consider the second optimization. In some cases, we can use the Word-Level index in the second search. If the first search provides a small amount of results, for example, less than 15, then this optimization can be used. In this case, we reinforce the second search in the following manner.

1) We exclude only the stop lemmas from $Q$.

2) We use the Word-Level index instead of the Document-Level index. That means that we process all posting records for non-excluded queried lemmas.

Our experiments show the following. Every query of the test query set was evaluated, and the two-step search process was used. Let us consider the queries in the evaluation process in which, the second optimization was used. The total time of the second search in these evaluations equals 14.8\% of the total search time (the total search time includes both steps). The total time of the second search when all test queries are considered equals 39\% of the total search time.

\sect{\label{t-search-results}Experiments}

We used the well-known GOV2 text collection. The total size of the collection is approximately 426 GB. The text collection consists of 25 million documents. If HTML tags are removed, then there is approximately 167 GB of plain text. The average text length of the documents is approximately 7 KB. 

The test query test contains the following query sets:
\begin{itemize}
\item	title queries from the TREC Robust Task 2004 (250 queries in total), 
\item	title queries from the TREC Terabyte Task from 2004 to 2006 (150 queries in total), 
\item	title queries from the TREC Web Task from 2009 to 2014 (300 in total), and
\item	title queries from the TREC Terabyte Task 2006 Efficiency Topics (10000 queries in total). 
\end{itemize}

Let us remove duplicates. The final test query set consists of 10690 queries.

We created two indexes as follows.

1) The ordinary inverted index.

2) Our additional indexes, which include the ordinary inverted index with the $NSW$ records and the $(w, v)$ and $(f, s, t)$ indexes, where $MaxDistance$ = 12.

In \cite{Jansen:2000:RLR:342495.342498}, logs of a full-text search system were analyzed. The following was shown. 

\begin{itemize}
\item	The users examined 2.35 search result pages on average.
\item	Many users, 58\% of them, examined only one page.
\item	Most users, 86\% of them, examined no more than three pages.
\item	One page contained 10 search results.
\end{itemize}

The number of users that examined only one page increased over time \cite{NineSearchEngineTL}. At some point, this number became 73\% of all users for US search systems. 

Moreover, in \cite{Jansen:2000:RLR:342495.342498}, the lengths of the search queries were analyzed. The following results are shown.

\begin{itemize}
\item	The average query length was 2.21 in words.
\item	Queries whose length is greater than 5 are very rare.
\item	The number of queries, whose length is 6, is approximately equal to 1\% of the total number of queries.
\item	Fewer than 4\% of all queries had lengths greater than 6.
\item	The number of queries that consist of one to three words is approximately equal to 80\% of the total number of queries.
\end{itemize}

Therefore, we limit our analysis to the first three pages of the search results (the first 30 search results). We also carefully consider the first page of the search results. The test query set contains the following.

\begin{itemize}
\item	The number of queries whose length is less than or equal to 3 is 4710.
\item	The number of queries whose length is less than or equal to 5 is 8788.
\end{itemize}

In Table \ref{tab:VeretennikovA-L-10-30}, the average values of $Precision$, the Levenshtein distance and $NDCG$ are shown in relation to the selected relevance functions. To calculate the metrics for a specific query, first the first 10 search results were taken into account, and then the first 30 search results were used.

The first three columns of the table correspond to cases when the first 10 search results were used. The first column contains values that were calculated based on the queries whose length is less than or equal to 3. For the second column, 5 was used as the limit of the query length. For the third column, 9 was used.
 
The following three columns of the table correspond to the cases where the first 30 search results were used. For the first of these columns, 3 is the limit of the query length, and 5 and 9 are the limits for the second and third columns.

Therefore, we show the results for the queries that contain no more than 3 words in a separated column. This is an important case according to \cite{Jansen:2000:RLR:342495.342498}. For this case, the search with additional indexes shows good results for every relevance function. For $R_{WeiSum, 0, 0.1, 0.9}$, we have $NDCG@10 = 0.980$. For $R_{TP,\text{BM25}}$, $R_{TP,\text{TF-IDF}}$, $R_{WeiSum, 0, 0.5, 0.5}$, $R_{WeiSum, 0, 0.25, 0.75}$, $R_{WeiSum, 0, 0.1, 0.9}$, and $R_{IntervalSumSq}$, the $NDCG@10$ values are greater than 0.96. When the length of queries increases, the values of the metrics decrease insignificantly.

\begin{table}[h!tbp]
  \caption{The results of the experiments, $L = |Q|$}
  \setlength{\abovecaptionskip}{1pt}
  \setlength{\belowcaptionskip}{-10pt}
  \setlength{\abovedisplayskip}{0pt}
  \setlength{\belowdisplayskip}{0pt}
  \label{tab:VeretennikovA-L-10-30}
\begin{tabular}{|l|l|l|l|l|l|l|}
\multicolumn{7}{c}{The average value of the Levenshtein distance} \\
\hline
Функция & 10, $L \leq 3$ & 10, $L \leq 5$ & 10, $L \leq 9$ & 30, $L \leq 3$ & 30, $L \leq 5$ & 30, $L \leq 9$\\
\hline
$R_{TP,\text{BM25}}$ & 0.542 & 1.308 & 1.621 & 2.264 & 4.701 & 5.474 \\ 
\hline
$R_{TP,\text{TF-IDF}}$ & 0.497 & 1.264 & 1.571 & 2.096 & 4.527 & 5.283 \\ 
\hline
$R_{WeiSum, 0, 0.75, 0.25}$ & 1.298 & 1.863 & 1.964 & 5.042 & 6.615 & 6.765 \\ 
\hline
$R_{WeiSum, 0, 0.5, 0.5}$ & 1.039 & 1.585 & 1.722 & 4.162 & 5.773 & 6.039 \\ 
\hline
$R_{WeiSum, 0, 0.25, 0.75}$ & 0.847 & 1.373 & 1.535 & 3.588 & 5.190 & 5.550 \\ 
\hline
$R_{WeiSum, 0, 0.1, 0.9}$ & 0.751 & 1.262 & 1.449 & 3.293 & 4.955 & 5.410 \\ 
\hline
$R_{IntervalSum}$ & 2.929 & 4.562 & 5.150 & 10.480 & 15.636 & 17.215 \\ 
\hline
$R_{IntervalOpt}$ & 1.873 & 2.885 & 3.042 & 7.142 & 9.823 & 10.068 \\ 
\hline
$R_{IntervalSumSq}$ & 1.352 & 2.325 & 2.579 & 5.817 & 8.576 & 8.989 \\ 
\hline

\multicolumn{7}{c}{The average value of $Precision$} \\

\hline
$R_{TP,\text{BM25}}$ & 0.968 & 0.909 & 0.887 & 0.958 & 0.900 & 0.886 \\ 
\hline
$R_{TP,\text{TF-IDF}}$ & 0.971 & 0.913 & 0.892 & 0.960 & 0.904 & 0.890 \\ 
\hline
$R_{WeiSum, 0, 0.75, 0.25}$ & 0.917 & 0.875 & 0.870 & 0.903 & 0.867 & 0.867 \\ 
\hline
$R_{WeiSum, 0, 0.5, 0.5}$ & 0.938 & 0.898 & 0.891 & 0.925 & 0.890 & 0.886 \\ 
\hline
$R_{WeiSum, 0, 0.25, 0.75}$ & 0.954 & 0.917 & 0.908 & 0.941 & 0.908 & 0.902 \\ 
\hline
$R_{WeiSum, 0, 0.1, 0.9}$ & 0.962 & 0.929 & 0.918 & 0.951 & 0.918 & 0.912 \\ 
\hline
$R_{IntervalSum}$ & 0.771 & 0.644 & 0.595 & 0.765 & 0.637 & 0.593 \\ 
\hline
$R_{IntervalOpt}$ & 0.894 & 0.826 & 0.815 & 0.882 & 0.833 & 0.829 \\ 
\hline
$R_{IntervalSumSq}$ & 0.919 & 0.845 & 0.826 & 0.895 & 0.828 & 0.818 \\ 
\hline

\multicolumn{7}{c}{The average value of $NDCG$} \\

\hline
$R_{TP,\text{BM25}}$ & 0.978 & 0.932 & 0.911 & 0.976 & 0.927 & 0.906 \\ 
\hline
$R_{TP,\text{TF-IDF}}$ & 0.981 & 0.936 & 0.915 & 0.978 & 0.930 & 0.910 \\ 
\hline
$R_{WeiSum, 0, 0.75, 0.25}$ & 0.950 & 0.917 & 0.911 & 0.919 & 0.884 & 0.882 \\ 
\hline
$R_{WeiSum, 0, 0.5, 0.5}$ & 0.964 & 0.933 & 0.927 & 0.942 & 0.907 & 0.903 \\ 
\hline
$R_{WeiSum, 0, 0.25, 0.75}$ & 0.973 & 0.948 & 0.941 & 0.958 & 0.927 & 0.921 \\ 
\hline
$R_{WeiSum, 0, 0.1, 0.9}$ & 0.980 & 0.959 & 0.951 & 0.968 & 0.941 & 0.933 \\ 
\hline
$R_{IntervalSum}$ & 0.826 & 0.730 & 0.684 & 0.800 & 0.696 & 0.653 \\ 
\hline
$R_{IntervalOpt}$ & 0.948 & 0.906 & 0.896 & 0.909 & 0.860 & 0.853 \\ 
\hline
$R_{IntervalSumSq}$ & 0.965 & 0.904 & 0.882 & 0.934 & 0.867 & 0.847 \\ 
\hline
\end{tabular}
\end{table}

The $NDCG$ values are closer to 1 than $Precision$. $NDCG$ is a better metric than $Precision$ for at least two reasons. The first reason is the following: to calculate $NDCG$, the values of the relevance scores need to be taken into account. A document that occurs at the start of the search result list is more important than a document that occurs at the end of this list. This is also considered when $NDCG$ is calculated. Second, if $NDCG$ is greater than $Precision$, then the following can be the reason for it: if some search results were missed in the search with our additional indexes in contrast to the ordinary search, then these search results are probably low relevant.

For the $R_{TP,\text{BM25}}$, $R_{TP,\text{TF-IDF}}$, $R_{WeiSum, 0, 0.1, 0.9}$, and $R_{IntervalSumSq}$  metrics, the results are the most promising, regardless of the query length. When the $R_{IntervalSum}$ metric is considered, the weight of an interval that contains the queried words is inversely proportional to the length of this interval. The results for the $R_{IntervalSum}$ metric are significantly worse than the results for metrics, in which the weight of the interval is inversely proportional to the square of the length of the interval. Most likely, when our additional indexes are used, a square dependence should be employed. Accordingly \cite{ProximityWithPair}, the square dependence is used for the majority of modern methods.

When the weighted sum methods $R_{WeiSum, 0, x, y}$ are considered, the metrics show better values when the weight of $TP$ is increased. Moreover, $NDCG@10$ is greater than 0.95 when $R_{WeiSum, 0, 0.1, 0.9}$ is used, regardless of the query length. Most likely, the value of $TP$ decreases very fast when the length of the interval increases and the weight of the interval is inversely proportional to the square of the length of the interval. Perhaps we should adopt the approach from \cite{TermProximityPerspective}. In \cite{TermProximityPerspective}, some floating-point parameter $\gamma \in [1,2]$ was introduced. The weight of the interval can then be defined as inversely proportional to the length of the interval raised to the power of $\gamma$.

Let us consider the $R_{IntervalOpt}$ metric. In this case, the metric values decrease significantly when longer queries are considered. For example, $NDCG@10 = 0.948$ when queries that contain no more than three words are considered. However, $NDCG@10 = 0.896$ when queries with lengths up to 9 are considered.

\begin{figure*}[tbp]
\setlength{\abovecaptionskip}{1pt}
\setlength{\belowcaptionskip}{-10pt}
\setlength{\abovedisplayskip}{0pt}
\setlength{\belowdisplayskip}{0pt}
\begin{center}
\begin{psfrags}
\small
\includegraphics[width=450pt,height=200pt]{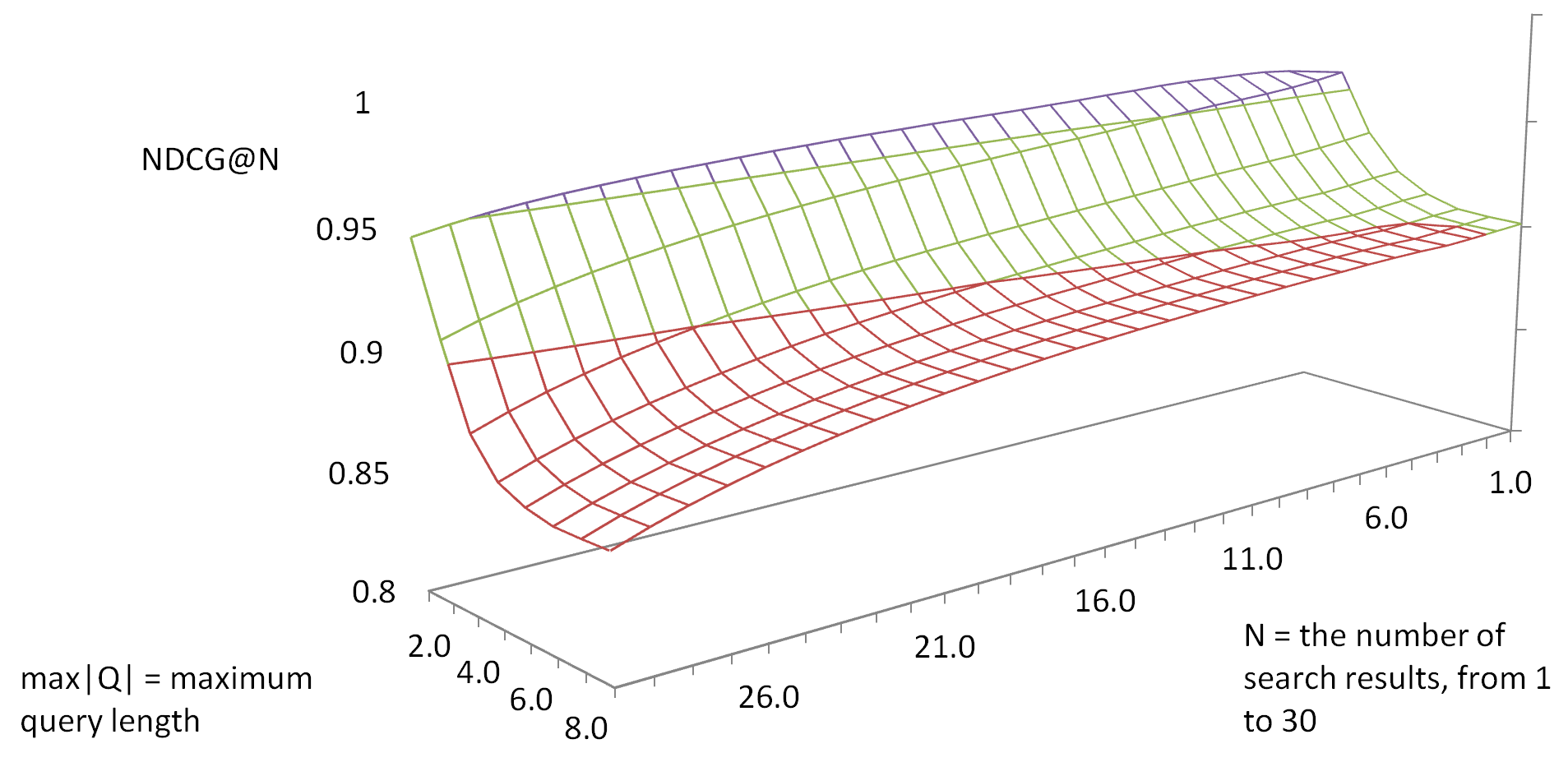}
\end{psfrags}
\end{center}
\caption{The $NDCG@N$ values for $R_{IntervalOpt}$}\label{VeretennikovA-Image-IntervalOpt}
\end{figure*}

In Fig.\,\ref{VeretennikovA-Image-IntervalOpt}, we show $NDCG@N$ for the $R_{IntervalOpt}$ metric. 

\begin{figure*}[tbp]
\setlength{\abovecaptionskip}{1pt}
\setlength{\belowcaptionskip}{-10pt}
\setlength{\abovedisplayskip}{0pt}
\setlength{\belowdisplayskip}{0pt}
\begin{center}
\begin{psfrags}
\small
\includegraphics[width=450pt,height=200pt]{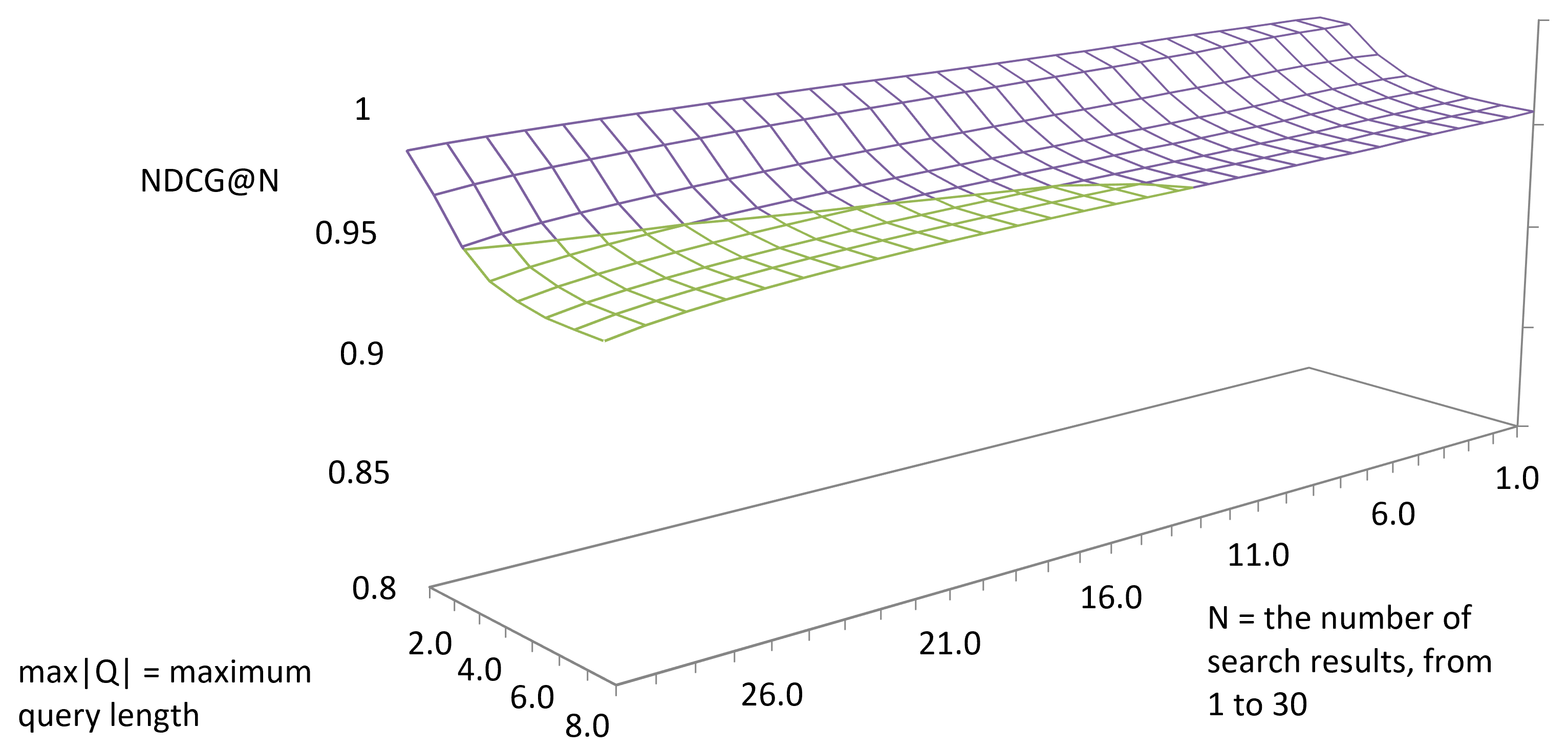}
\end{psfrags}
\end{center}
\caption{The $NDCG@N$ values for $R_{WeiSum, 0, 0.1, 0.9}$}\label{VeretennikovA-Image-WeiSum}
\end{figure*}

In Fig.\,\ref{VeretennikovA-Image-WeiSum}, we show $NDCG@N$ for the $R_{WeiSum, 0, 0.1, 0.9}$ metric.

On the $max|Q|$ axis, we plot the maximum query length for the queries, which are used to calculate $NDCG@N$. On the $N$ axis, we plot $N$, that is, the number of search results, which are used to calculate $NDCG@N$. The $NDCG@N$ value is shown in the $0.8$ to $1$ range.
 
When $R_{IntervalOpt}$ is considered we have the following. The $NDCG@N$ value decreases significantly when the $max|Q|$ value is increased. However, the $NDCG@N$ value does not change significantly when $N$ is changed.

When $R_{WeiSum, 0, 0.1, 0.9}$ is considered we have the following.  The $NDCG@N$ value decreases insignificantly when the $max|Q|$ value is changed.

We compare these two metrics for the following reasons. When $R_{IntervalSum}$ is used, the results are the worst. Here, the weight of an interval is inversely proportional to the length of the interval. This approach has little application when our additional indexes are used. For $R_{IntervalOpt}$, we have slightly better results and this method can have some application. From an applicability point of view, $R_{IntervalOpt}$, directly follows $R_{IntervalSum}$. When $R_{WeiSum, 0, 0.1, 0.9}$ is considered, the results are the best. Essentially, we compare the worst and the best methods, which can be employed when our additional indexes are used.

\sect{\label{t-conclusion}Conclusion}

A methodology to evaluate the search quality when additional indexes are employed was presented. We considered several relevance calculation methods and established the following: for several relevance calculation methods, the search results obtained in the search with our additional indexes and the search results obtained in the search with ordinary inverted indexes are similar in terms of relevance. When $R_{TP,\text{BM25}}$, $R_{TP,\text{TF-IDF}}$, $R_{WeiSum, 0, 0.1, 0.9}$, and $R_{IntervalSumSq}$ metrics are used, the best results are achieved. 

Let us consider $R_{WeiSum, 0, 0.1, 0.9}$. When the first ten search results are taken into account, the $NDCG$ value is larger than 0.95. The maximum possible value of $NDCG$ is 1. The $NDCG$ metric is well-known and commonly used when the search quality is analyzed.

The two-step search method was enhanced to improve the search quality. First, the proximity search is performed with our additional indexes. Second, the non-proximity or partially-proximity search is performed with the ordinary index. The total time of both of these steps is significantly less than the total time of the proximity search in the ordinary index.

Future research can be performed in the following directions. The $R_{IntervalOpt}$ function is calculated with the following approximation. Only intervals that contain all queried words are taken into account. An algorithm that considers intervals that contain some subset of the queried words can be useful. In our work, the set of all possible queries is divided into several classes, namely QT1-QT5. Different algorithms were developed to process the different classes or types of queries \cite{ProximityFTWithRTG, ProximityFTMultiComponentKeys}. We spent a significant amount of time unifying the relevance calculation process for these different algorithms. It is important to develop a unified search algorithm. In addition, it is interesting to consider other relevance calculation methods and relevance functions.

\pagestyle{basestyleeng}

\selectlanguage{english}

\renewcommand{\refname}{{\rm

\receivedeng

\contactseng

\vspace{16pt}

See also:

\href{http://www.veretennikov.ru/}{http://www.veretennikov.ru/}

\href{http://www.veretennikov.org/Default.aspx?f=Publish\%2fDefault.aspx\&language=en}{http://www.veretennikov.org/Default.aspx?f=Publish\%2fDefault.aspx\&language=en}

\end{document}